# An Improved AIS Based E-mail Classification Technique for Spam Detection

**Ismaila Idris**
Dept of Cyber Security Science, Fed. Uni. Of Tech. Minna, Niger State
Idris.ismaila95@gmail.com

**Abdulhamid Shafi'i Muhammad**
Dept of Cyber Security Science, Fed. Uni. Of Tech. Minna, Niger State
Shafzon@yahool.com

*Abstract* - An improved e-mail classification method based on Artificial Immune System is proposed in this paper to develop an immune based system by using the immune learning, immune memory in solving complex problems in spam detection. An optimized technique for e-mail classification is accomplished by distinguishing the characteristics of spam and non-spam that is been acquired from trained data set. These extracted features of spam and non-spam are then combined to make a single detector, therefore reducing the false rate. (Non-spam that were wrongly classified as spam). Effectiveness of our technique in decreasing the false rate shall be demonstrated by the result that will be acquired.

*Keywords* - Algorithm, Artificial immune system, E-mail Classification, Non-Spam, Spam

## I. INTRODUCTION

Inspiration provides biological system in developing algorithm and methods. Algorithm developers are actively looking for a breakthrough in immune system. Models and application in artificial immune system is coming up as an active and attractive field of great diversity. Great source of inspiration to computational model are drawn from the already knowledge of the immune system. Classification is one of the familiar techniques used in machine learning. Patterns belonging to different classes are discriminated due to the generation of decision boundaries, It is then divided in to training set and testing set randomly and then classification is made on the training set were as the testing set is used to assess performance of the generated classifier. Spam is often sent as a commercial content. Spam is defined as unsolicited e-mail [1] Spam is a very important subject that most be prevented by both technologist and the law of the land. In distinguishing spam messages, various techniques has been proposed by researchers in other to fight against e-mail spam whose results are not very effective due to constant change in patterns of the spam behavior. [2]. Over the past years, rapid expansion of computer network system as change the world. It is essential for an effective computer security system because attacks and criminal intend are increasingly popular in computer network. Golovko et al. [3] . There are several measure put in place by many companies in the area of creating anti spam software based on signatures and have a very efficient performance in detecting spam fast. Though, new variation of spam and unknown spam are very difficult to detect by this software. The traditional way of detecting spam based on signature is no more efficient for today systems. Spam detection and e-mail classification problem are been solved using Artificial Immune System [4]. An improved e-mail classification technique based on Artificial Immune System shall be designed and implemented. We shall first of all generate a spam and non-spam detector after which email classification will take place by utilizing the non-spam and the spam accordingly in other to successfully reduce the false rate. Our improved classification techniques are also compared with the existing techniques. The experiment confirms the reliability and efficiency of our new techniques in minimizing false positives. The datasets used in this research is gotten from machine learning repository, Center for Machine Learning and Intelligent System.

The organization of the remaining part of this paper is as follows: a theoretical presentation of the background of related work in Artificial Immune System in general was presented in section 2. Classification techniques based on Artificial Immune System was discussed in section 3. Section 4 presents the Experimental frame work of our improved techniques. Results was presented and analyzed in section 5. Highlight on direction of future work with conclusion was finally discussed in section6.

## II. RELATED WORK

Artificial Immune System inspired by biological Immune System. [5] is an emerging learning techniques. Most of the classification algorithms are used for solving spam problems and most of the techniques focus on machine learning in spam filtering, this learning techniques are: Rule Learning, Naïve Bayes, Support Vector Machine, Artificial Neural Network, Artificial Immune System (AIS), DNA Computing, decision trees and combinations of different learning. During these years, several classifiers such as naïve Bayes, text compression and artificial neural network have been proposed to detecting and handling the spam. These classifiers are based on probabilistic techniques and machine learning. The following paragraphs represent the related researches summarily.

[6] used the SVM (support vector machines) algorithm for classifying the emails as spam or non-spam. In addition, besides this algorithm they used and compare it to the three other classification algorithms: Ripper, Rocchio and Boosting decision trees. They apply these algorithms on two different data sets: one of them constrained to 1000 best features and the other one constrained over 700 features. In terms of accuracy and speed, Boosting trees and SVM had the acceptable performance. But, compare to the other three algorithms, SVM's result shows the less training time against the other algorithms.

Biological immune system is created to give support and protection to the body against antigens. Its ability for recognition enables it to be able to differentiate between antigens (non-spam) and the body cell (self). Leandro et al. characterized the immune system by its recognition of foreign body, noise tolerance and uniqueness [5]. The emergence of artificial immune system was since the last decade; Inspired by biological immune system, it is been incited by researchers to create immune based models for computer security system. Different AIS techniques and model are been research on and also literatures since the work of Dasgupta et al. [6].This models are mainly, negative selection mechanism, Clonal selection rules and immune network theory.

Several artificial immune algorithm are been created with imitation of Clonal selection theory. The Clonal selection principle comprises of antigen recognition and differentiation in the memory cell. Burnet [7] proposed the theory and adaptive immune system to antigen stimulus basic responses was elaborately explained with the theory. A Clonal selection algorithm called CLONALG was then proposed by Leandro et al. [8] for optimization and learning, an N number population of antibody is been generated by CLONALG, a random solution is been specified for the optimization process. Selection, cloned and mutation takes place through the selection of some of the good antibody during iteration session. This resulted into new antibody, were the best among them merges with the original population while worst antibody of the previous generation are been substituted with randomly generated new ones.

The ability of the immune system been utilized to detect antigen that are not known to respond to self cells is the Negative selection mechanism; while protecting the body over self reacting lymphocytes. Receptors are created through pseudorandom genetic rearrangement procedure during the generation of T-cell, which then go through a censoring procedure in the thymus; this process is known as negative selection. With this procedure, response against self protein by T-cells is destroyed while only those that were able to match to self protein are given the chance to leave the thymus. This T-cell that are allowed is then lunched in to the body protecting it against antigens. [9]. Negative selection algorithm was proposed by Forrest et al. [10]. The principle behind it is to create a set of detectors that could be use to detect malware. This was achieved by randomly creating candidates and those that recognize training spam data are been discarded. In furtherance of our related work, we are going to give a summary of some of the existing work of the different AIS techniques and models of the previous five years in this section

Perfect feature sequence and multiple features consider by Bayes classifiers, but other classifiers usually use pruning and text pre-processing [11]. Generally, Bayes-based techniques are well known to achieve high spam detection accuracy either as standalone classifiers or as parts of classifier ensembles. [12] Used feature extraction for spam detection. Basically, they used AIS (Artificial Immune System) for feature extraction. This method extracts a comparatively small set of features which are used as inputs in classification to spam detection. These features are modeled by regular expressions of terms. Features are created from the content of spam messages by using the strings and character matching rules. One of the algorithm that are mentioned and compared with them in their research as the spam detection model is a Back propagation neural network. None of the anti spam solution that has been proposed on false positive and false negative approach perfection [13]. Though the result of spam is reduced but not completely. More so, the false positive is more problematic and important than the false negative. Therefore, most researchers and developers are trying to completely get rid of possible false positive mistakes.

## III. CLASSIFICATION TECHNIQUES BASED ARTIFICIAL IMMUNE SYSTEM

E-mail classification exists with two fundamental problems by existing anti-spam method. It is either the email is recognized as spam and is deleted or non-spam and been accepted carelessly. This process is called false positive and false negative respectively. The false positive occurs when the email or data that are needed to create a detector are classified as spam while emails or data that are supposed to be discarded are recognized as non-spam. This scenario (false negative and false positive) is calculated using classification accuracy which is the main measure of performance. Classification accuracy deals with false positive, false negative and accuracy whose formulae are used to compare different classifier performance [14]. False positive is the percentage of non-spam data classified as spam while false negative is the percentage of spam data which are classified as nonspam and accuracy is calculated by the formulae below.

$$\text{Accuracy} = ((TP+TN)/(S+NS)) \times 100 \qquad (1)$$

Where TP represent True Positive; TN represent True Negative; S represent Spam and NS represent Non Spam.

The figure below demonstrates how false positive and false negative are calculated. The first row depict the total non-spam that is divided to true positive and false positive. These rows contain total dataset which are non spam and some are wrongly classified as spam (FP) while others are assigned correctly as non-spam while the opposite is the case with the second row.

TABLE 1
FALSE POSITIVE AND FALSE NEGATIVE

|  | Non-spam | spam |
|---|---|---|
|  | True positive (TP) | False negative (FP) |
|  | False negative (FN) | True negative (TN) |

None of the anti spam solution that has been proposed on false positive and false negative approach perfection [13]. Though the result of spam is reduced but not completely. More so, the false positive is more problematic and important than the false negative. Therefore, most researchers and developers are trying to completely get rid of possible false positive mistakes. The processes of classification based on Artificial Immune system are in stages and are summarized as follows:

Our classification techniques comprises of the pre-processing face which encompasses the transformation code, e-mail vector extractor, text body and header separation. The best way we reflect the message characteristics is through extraction which is the process of choosing a vector set. [12]. The size of attachment and text, the word in message and also HTML code are some of the component of the vector. This component varies in importance and can represent the email's characteristics.

The Training generates set of self detector due to most represented self. Best suitable detector can be realized after experiment by adjusting the value at all time. This is due to the center and radius present in the detector. Newer e-mails are been compare with existing detector in other to decide if to add the new e-mails to the detector or not. Though, it is disregarded if the self is closer to the center of any of the detector. This is so as to allow the detector to have a wider range of coverage area instead of having much aggregation.

Also, as a result of the indeterminate length of the e-mail vector, it becomes a tax calculating affinity of two variable length vectors which is defined as below.

$$\text{Affinity}(x,y) = \sum match(x,y)/l, / x /= l \quad (2)$$

x = Affinity of e-mail x
y = Affinity of e-mail y
L = Denote shortest length of x and y
n and I ensures the affinity is between the range 0 or 1

$$\text{Match}(x,y) = \begin{cases} 1, & \sum_{i=0}^{n} match(x \to x_i, y)/l, / x /= l \\ 0, & otherwise \end{cases} \quad (3)$$

$$\text{Match}(x,y) = \begin{cases} 1, & \sum_{i=0}^{n} match(y \to y_i, x)/l, / y /= l \\ 0, & otherwise \end{cases} \quad (4)$$

Where, the Matching of x is the total match value between memory-spam x And the Matching of y is the total match value between memory- non-spam y

The dataset is given as:

$$S = \{<x, y, \text{affinity}> | x, y \in \cup_{i-1}^{n} H, \text{affinity} \in N\}a \quad (5)$$

Affinity is the ability of the spam to match with both spam and non-spam. This exists when memory-spam and memory-non-spam are generated. The set is divided in to spam training set and non-spam training set. The trained detectors is used to classify in coming e-mail by obtaining feature vector after pre-processing when a new e-mail arrives and both e-mail and detectors affinity are calculated. When we have affinity that is greater than threshold, it is said to be spam; otherwise it's a non-spam. Then we have intervention manually by the users retraining and error correction. With these, the accuracy and early timing of the detector is certain. In our optimized e-mail classification method, evaluating the classification method is best achieved by the reduction of false rate. This aspect of reducing the false rate of e-mail classification is a vital aspect of our research work.

Therefore, this research not only depends on the recall rate. A new classification method is been design in this paper with the use of two detector set. The spam and non-spam detector. They are combined together to form an effective detector thereby reducing false rate were spam is considered as nonspam and then discarded.

The proposed method consists of suspicious spam extractor and a suspicious spam detector as shown above. In the suspicious spam extractor; both the self detector set and the non-self detector set are generated from the dataset, this is taking from the spam and non-spam programs of the training data set with the help of the suspicious spam extractor. After securing suspicious program (self and non-self) in the suspicious self detector using the suspicious self extractor, an r matching rules will be initiated and computed between the suspicious program and the memory self and memory non-self will be established resulting in to a new memory detector.

The figure below illustrates set of self and non-self detectors that are generated during the training procedure.

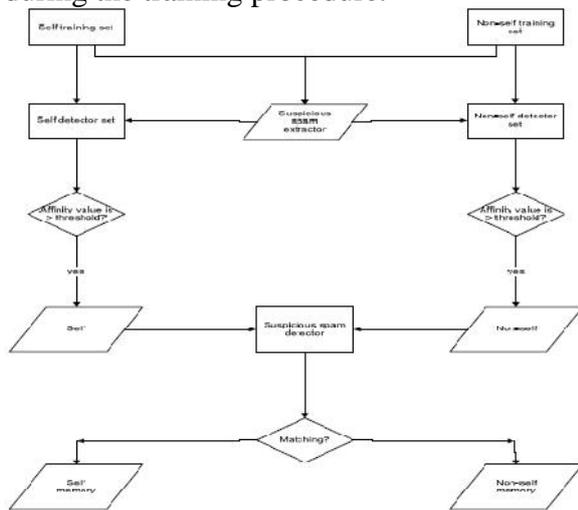

**Fig.1** Improved E-Mail Classification Techniques

### IV. EXPERIMENTS

We use the machine learning repository from the center for machine learning and intelligent system for classifying e-mail as self and non-self. The 'spam base' last column indicate if the e-mail was considered spam (1) or non-spam (0). The data set used in these techniques has 4601 instances in which 39.4% are spams and each of the instances has 57 attributes. This data set was divided into two classes. We have the training data set and the testing data set divided in the ratio 60% to 40%. From the training data set, the self and non undergo a preprocessing stage, finally generating about 100 self and non detectors. We select an e-mail from the training data set for word segmentation, and generate a set of self detector by training, which consist of word list, number of detected e-mail and number of matched e mail. As well we calculate the affinity on property wordlist. If affinity is greater than threshold, we add one to the number of matched e-mail. Also number of e already detected will be irrespective whether the affinity is less than or greater than the threshold.

Consequently, a set of non-spam generated similar to that of s We select an e-mail from the training data set for word segmentation. In case affinity, the object to be compared is the spam detector that was earlier generated instead of the non-spam. If any value is said to be greater than threshold, we add the generated detector to the non set. We compare a non-spam spam detector in other to find detectors which are similar to spam but rather belong to non-spam. Other experiment for testing by using the trained detectors generated are carried out after training

### V. RESULTS AND ANALYSIS

Result obtained from the im classification techniques and the formal techniques are represented as below.

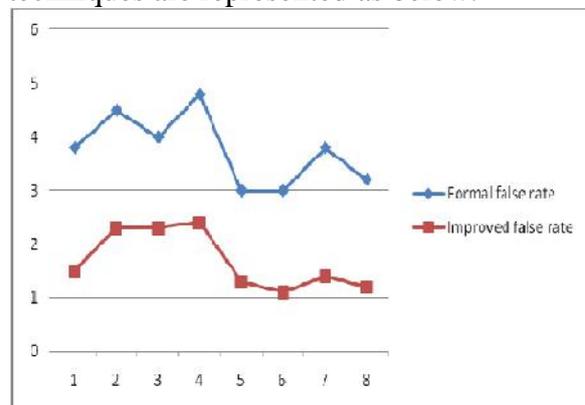

**Fig.2** Classification of False Rate

The result of the traditional (formal) false rate against our improved classification is shown in figure 2 above. For our improved false rate classification, our best rate is 1.2%, average is at 1.6% while the worst false rate is at 2.4%. The best rate for the existing techniques is at 3.0% while its average is at 3.8% and the worst false rate is 4.8%. We see how figure 2 analyses performance of our improved techniques comparing the improved technique and the formal technique. The improved classification is lower than the existing classification, therefore, reducing the false positive rate and improving the spam detector.

## VI. CONCLUSION

An improved e-mail classification techniques based on Artificial Immune System proposed in this paper, the essence is to reduce false positive and create spam detectors. We make use of spam and non-spam in our training data set. This process is very effective in reducing the false rate but its drawback is the reduction of the recall rate. Our subsequent research work will be looking at how to rate and the false rate by making appropriate tradeoff between them.